\documentclass[times,twocolumn,final]{elsarticle}
\usepackage{medima}
\usepackage{framed,multirow}

\usepackage{amssymb}
\usepackage{latexsym}
\usepackage{booktabs}
\usepackage{makecell}
\usepackage{multirow}
\usepackage{graphicx}
\usepackage{amsmath}
\usepackage{comment}
\usepackage[ruled,linesnumbered]{algorithm2e}
\newcommand{\rev}[1]{{\textcolor{black}{#1}}}
\newcommand{\revv}[1]{{\textcolor{black}{#1}}}

\usepackage{url}
\usepackage{xcolor}

\usepackage{hyperref}

\definecolor{newcolor}{rgb}{.8,.349,.1}

\journal{Medical Image Analysis}

\begin{document}
\verso{Wang \textit{et~al.}}

\begin{frontmatter}

\title{Real-time landmark detection for precise endoscopic submucosal dissection via shape-aware relation network}

\author[1]{Jiacheng Wang \fnref{fn1}}
\author[3]{Yueming Jin \fnref{fn1}}
\fntext[fn1]{They contribute equally to this work.}
\author[4]{Shuntian~Cai}
\author[4]{Hongzhi~Xu}
\author[3]{Pheng-Ann~Heng}
\author[6]{Jing~Qin}
\author[1]{Liansheng Wang\corref{cor1}}
\ead{lswang@xmu.edu.cn}
\cortext[cor1]{Corresponding author.}

\address[1]{Department of Computer Science at School of Informatics, Xiamen University, Xiamen 361005, China}
\address[3]{Department of Computer Science and Engineering, The Chinese University of Hong Kong, China}
\address[4]{Department of Gastroenterology, Zhongshan Hospital affiliated to Xiamen University, Xiamen, China}
\address[6]{Center for Smart Health, School of Nursing, The Hong Kong Polytechnic University, Hong Kong}

\begin{abstract}
We propose a novel shape-aware relation network for accurate and real-time landmark detection in endoscopic submucosal dissection (ESD) surgery.
This task is of great clinical significance but extremely challenging due to bleeding, lighting reflection, and motion blur in the complicated surgical environment. 
Compared with existing solutions, which either neglect geometric relationships among targeting objects or capture the relationships by using complicated aggregation schemes, the proposed network is capable of achieving satisfactory accuracy while maintaining real-time performance by taking full advantage of the spatial relations among landmarks.   
We first devise an algorithm to automatically generate relation keypoint heatmaps, which are able to intuitively represent the prior knowledge of spatial relations among landmarks without using any extra manual annotation efforts.
We then develop two complementary regularization schemes to progressively incorporate the prior knowledge into the training process.
While one scheme introduces pixel-level regularization by multi-task learning, the other integrates global-level regularization by harnessing a newly designed grouped consistency evaluator, which adds relation constraints to the proposed network in an adversarial manner.
Both schemes are beneficial to the model in training, and can be readily unloaded in inference to achieve real-time detection.
We establish a large in-house dataset of ESD surgery for esophageal cancer to validate the effectiveness of our proposed method.
Extensive experimental results demonstrate that our approach outperforms state-of-the-art methods in terms of accuracy and efficiency, achieving better detection results faster.
Promising results on two downstream applications further corroborate the great potential of our method in ESD clinical practice.
\end{abstract}

\begin{keyword}

\KWD \\ Endoscopic submucosal dissection \\ Landmark detection \\ Shape-aware relation network \\ Real-time detection
\end{keyword}

\end{frontmatter}


\section{Introduction}
\begin{figure*}[!t]
\includegraphics[width=\textwidth]{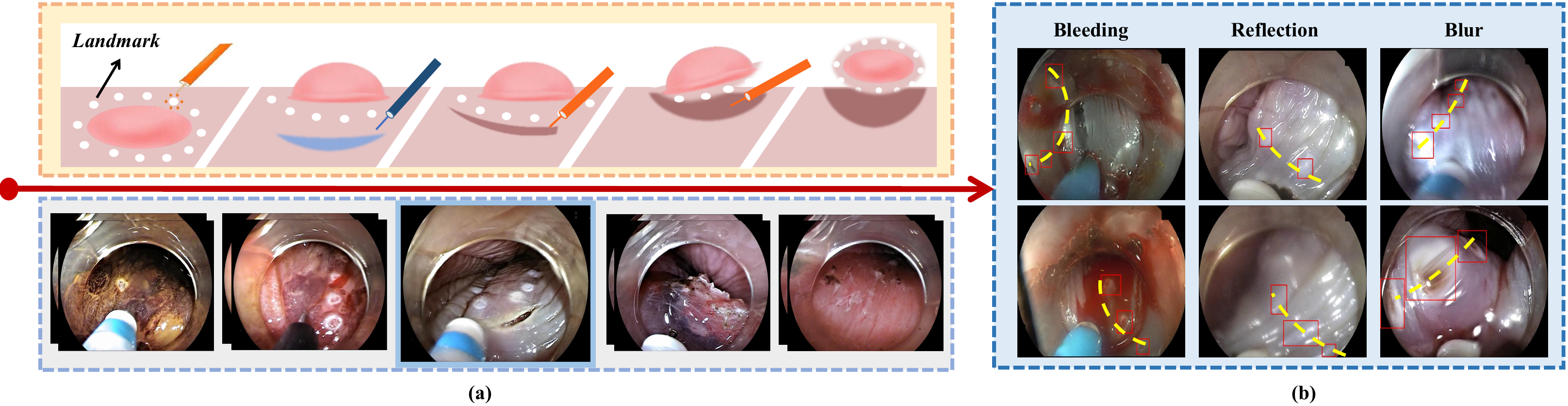}
\caption{(a) Detailed steps of ESD surgery. The first row is animation diagram and the second row is the related scenario in clinical practice. Highlighted blue section is the key step \emph{circumferential cutting} in ESD surgery, which is also the portion this work focused on. 
(b) Challenging cases in the \emph{circumferential cutting} step, due to bleeding, lighting reflection, and motion blur.
} \label{fig:challenges}
\end{figure*}
Endoscopic submucosal dissection (ESD) is a well-established technique of endoscopic resection for removal of epithelial lesions. Different from another endoscopic resection technique, i.e. endoscopic mucosal resection (EMR), which utilizes a snare or cap for lesion removal, ESD implements submucosal dissection using a knife. 
Therefore, ESD allows one-step and en bloc resection for larger and potentially deeper lesions~\citep{ono2016guidelines}. 
In addition, ESD is able to greatly alleviate the removal residual of lesions and reduce the risk of disease relapse. 
To the end, it is widely-adopted for lesion removal nowadays~\citep{oyama2005endoscopic,gotoda2006endoscopic,terheggen2017randomised}. 

As shown in Fig.~\ref{fig:challenges} (a), ESD generally consists of five steps: 
(1) creating landmarks around the lesion; the landmarks are used to label the boundary between the lesion and normal tissues as it is hard for the surgeon to well perceive the boundary during cutting,
(2) injecting saline to separate submucosal layer from muscle layer,
(3) cutting the mucosa surrounding the lesion according to the landmarks,
(4) submucosal dissection of the connective tissues under the lesion~\citep{shah2015endo,ono2016guidelines}, and
(5) hemostasis of cancer bleeding and post-operative assessment.
During the procedure, real-time perceiving the landmarks for precise cutting is a vital determinant for the success of an ESD surgery.

While landmarks are created for intuitively observing the boundary, it is challenging for surgeons to easily and consistently perceive the landmarks during cutting.
There are mainly three obstacles (see Fig.~\ref{fig:challenges}(b)).
 (1) Bleeding. Cutting of mucosa easily causes bleeding, and the blood probably obscures landmarks, making them unrecognizable. 
 (2) Lighting reflection. A water jet is usually used to irrigate the blood for improving the visibility of a surgical scene, yet leading to severe lighting reflection caused by water; it further increases the difficulty in identifying landmarks due to the extremely low intensity contrast with background tissues caused by the reflection.
 (3) Motion blur. Surgical scenes are often blurred due to the movement of endoscopic camera and background tissues, making it challenging to track landmarks.
Moreover, the frames with these obstacles generally account for a larger portion during the cutting in an ESD surgery when compared with other surgical procedures.
In this regard, real-time and accurate landmarks detection is highly demanded in ESD surgery to provide timely and precise guidance to surgeons.
\rev{However, considering the potential reason about the difficulty in data acquisition, there is very rare study about the computer-assisted ESD surgery.
The most related work is using convolutional neural network to estimate cancer invasion depth to help surgeons screen patients for endoscopic resection~\citep{zhu2019application}. }
There are, to our knowledge, no existing solutions for real-time and accurate landmarks detection for ESD surgery.

One straightforward thought is to harness existing surgical scene analysis algorithms for this task. 
However, these algorithms mainly focus on the tasks of lesion detection\rev{~\citep{cai2019using,kanayama2019gastric,ghatwary2019early,jha2021real}}, polyp detection in wireless capsule endoscopy\rev{~\citep{zhang2016automatic,yuan2019densely,xia2021use}} and colonoscopy video~\citep{bernal2017comparative,wang2018development,lee2020real}, as well as instrument detection in laparoscopic data\rev{~\citep{bouget2017vision,laina2017concurrent,allan20183,colleoni2019deep,yamazaki2020automated,islam2020ap}}.
These objects are quite different from landmarks in ESD surgical scenes, which have distinctive yet valuable geometric relationships \rev{that will greatly improve both accuracy and efficiency if they can be fully utilized.}    
In natural image processing domain, there exist some models that leverage relations among objects to facilitate the detection tasks.
However, most of them are dedicated to aggregating the features from object instances in order to augment the feature representations by using either non-local block architecture~\citep{Deng_2019_ICCV} or graph convolutional operations~\citep{xu2019spatial}.
Unfortunately, the feature fusion strategies often bring high computational costs, leading to low latency in inference.
To the end, they are not suitable for our task, in which real-time inference is of great importance, and the detection algorithm, we assume, should be deployed under a clinical setting with reasonable yet limited computational resources. 
Recently, while some real-time detectors have been proposed~\citep{law2018cornernet,zhou2019objects,Tan_2020_CVPR}, most of them did not take geometric relationships into account, and hence are insufficient to tackle the above-mentioned challenges.

In this paper, we present a novel shape-aware relation network for accurate and real-time landmarks detection in ESD surgery, aiming at enabling the surgeon to conduct precise lesion removal under complicated surgical environments.
We carefully observe the inherent geometric characteristics of these landmarks and decide to harness the shape/geometry constraints among them to tackle challenges caused by bleeding, lighting reflection, and motion blur, as, after all, most of these landmarks are located on the boundary (yellow dotted curve in Fig.~\ref{fig:challenges}(b)) between the lesion and normal tissues.  
To take full advantage of the shape/geometry constraints, we first design an algorithm to automatically generate a \emph{relation keypoint heatmap}, which indicates the middle point between each pair of neighbouring landmarks.
We further propose two complementary schemes, from \emph{pixel-level} and \emph{global-level} respectively, to progressively incorporate the shape regularization in the learning process.
We first develop a multi-task learning scheme to simultaneously detect the landmarks and the midpoints. 
Serving as an auxiliary task, locating pixel-wise midpoints is able to regularize the landmarks detection to make sure the results do not deviate from the boundary between the lesion and normal tissues.  
Then, we further propose a grouped consistency evaluator (GCE) in order to measure the spatial consistency of a landmark heatmap and a relation keypoint heatmap, after seeing a global-level view of the two maps.
The group of ground truths tends to achieve a higher score than the groups with both of or one of the heatmaps that are generated by the proposed network.
A consistency-based adversarial loss is thus designed based on the GCE to further drive the model to comply with the shape constraint.
More importantly, both of the two schemes are only being effective in the training processing, and will not be utilized in the inference; hence, they will not introduce extra computational costs and running time, maintaining the real-time performance of the proposed network.
In order to evaluate the proposed network, we conduct extensive experiments on a large in-house dataset.
Experimental results demonstrate that the proposed network outperforms state-of-the-art approaches in terms of accuracy and efficiency in the challenging task. 
Our contributions can be summarized as follows.

\begin{itemize}
    \item  We propose a novel shape-aware relation network for real-time and accurate landmarks detection to facilitate precise lesion removal in ESD surgery; the proposed network incorporates the inherent prior knowledge of spatial relations among landmarks as an additional regularization to tackle the challenges in complicated surgical environments.
    
    \item We propose two complementary regularization schemes to progressively integrate the spatial relations, one performing pixel-level regularization via multi-task learning and the other conducting global-level regularization via a grouped consistency evaluator. Both can be readily unloaded in inference to retain real-time performance.
    
    \item We construct a large dataset about ESD in esophageal lesion for this unexplored task with great clinical significance.
    Extensive experimental results show that our approach achieves superior performance over the state-of-the-art methods; we further demonstrate the potential of the proposed method in two downstream applications for precise ESD surgery.
    
\end{itemize}

\section{Related Work}
\subsection{Object Detection in Endoscopic Scene}
Object detection has witnessed plenty of successes in surgical scene analysis of endoscopy images. Most studies concentrate on the detection of lesion~\citep{ghatwary2019early}, polyp~\citep{lee2020real} or surgical instrument~\citep{yamazaki2020automated}. 
For lesion detection, \cite{ghatwary2019early} propose to adapt Faster R-CNN~\citep{fasterrcnn} to automatically identify regions of esophageal adenocarcinoma (EAC) from high-definition white light endoscopy (HD-WLE) images. It can help early detection and treatment of EAC while this network is computation-intensive, thus inappropriate for landmark detection. 
Polyp detection is also one of the most conventional detection tasks in endoscopy images. Among them, a recent study~\citep{lee2020real} employs a YOLOv2~\citep{redmon2017yolo9000} to develop the algorithm for automatic polyp detection, which has exhibited high detection sensitivity and rapid processing.
\rev{~\cite{LIU2021102052} propose a consolidated domain adaptive polyp detection framework to bridge the domain gap between different \revv{colonoscopic} datasets.} 
Nevertheless, landmark detection in ESD surgery is extremely harder than polyp detection as the surgical scenario after mucosal resection has more challenging cases and noise. 
As for instrument detection,~\cite{yamazaki2020automated} introduce the clinical employment of YOLOv3~\citep{redmon2018yolov3} for detecting surgical tools from laparoscopic gastrectomy video images. 
\rev{While, YOLOv3 still cannot handle landmark detection well as the landmarks show too similar semantics with neighbour tissue that they are hard to detect without extensive constraints.
Rethinking the difference between ESD landmark detection and previous studies, we take the first step to build our ESD detection model by leveraging its inherent characteristic, i.e., proposing the shape-aware regularization for landmark detection.}

\subsection{Efficient Object Detection}
With the increasing efficiency requirement in the practical application scenario, a detection network is highly demanded to keep satisfactory accuracy with retaining good efficiency.
In general, these methods have abandoned the design of Region Proposal Network~\citep{fasterrcnn} to save computation.
They could be grouped into two streams by using anchor or not.
Anchor-based methods represent each object through a series of axis-aligned anchors and use an additional classification model to determine if the image content is a specific object or background~\citep{redmon2018yolov3, lin2017focal, Tan_2020_CVPR}. These methods primarily provide a large amount of probable detection, thus is more likely to recognize all landmarks, yet suffer from too much false detection that may confuse surgeons. 
On the contrary, anchor-free detectors regard each object as a certain number of keypoints, such as a pair of top-left corner point and bottom-right corner point~\citep{law2018cornernet}, or four extreme points (top-most, left-most, bottom-most, right-most) with one center point ~\citep{zhou2019bottomup}. 
Most recently,~\cite{zhou2019objects} \revv{propose} to detect an object via detection of its center point and regression of its size and offset, the former of which can be regarded as key-point detection.
Treating an object as one point, has proved to be effective in plenty of downstream vision tasks such as video tracking~\citep{zhou2020tracking}, action recognition~\citep{li2020actions}, and interaction detection\citep{Wang2020IPNet}.
Nevertheless, simply employing this group is also insufficient for landmark detection, since methods in this group can not catch the shape-aware relation between landmarks, failing to handle aforementioned problems well.
Still, they provide a flexible and efficient foundation to improve landmark detection by exploring the shape-aware relation between landmarks.

\subsection{Relation Learning for Object Detection}
Before the prevalence of deep learning, involving the relation information between objects has been verified to help improve objects recognition in the computer vision community~\citep{5995720,torralba2004sharing}, 
and the effectiveness of leveraging relation between proposals in deep networks has also been proved in still image objection~\citep{Hu_2018_CVPR, xu2019spatial}, and video object detection~\citep{Deng_2019_ICCV,Wu_2019_ICCV,lin2020dual}. 
Most methods utilize the object relation by aggregating extracted features from other multiple object instances to generate enriched features for better object detection.
For example, \cite{Hu_2018_CVPR} \rev{design} relation distillation modules to measure the relation among proposals and then combines them to boost the feature representation for detection. 
\cite{xu2019spatial} \rev{present} a spatial-aware graph relation network which enables adaptive graph reasoning over proposals with an interpretable learned graph, to distill the contextual information and spatial information for better detection.
\cite{Deng_2019_ICCV} and \cite{Wu_2019_ICCV} propose to capture the long-range dependency of multiple instances in a video sequence. 
\cite{lin2020dual} \rev{demonstrate} that measuring the spatial similarity between instances is more reliable than appearance similarity, whose point is similar to \cite{xu2019spatial}. 

However, the aggregation would cause supernumerary computation, resulting in longer computing time, making these models hardly applied in clinical scenarios, especially for our ESD landmark detection task.
How to exploit the relationships to boost performance while retaining inference efficiency becomes the crucial yet challenging direction remaining to be explored.
In this work, we find a distinctive prior cue about the spatial relationship between landmarks.
Motivated by \cite{xu2019spatial}, in which an auxiliary task is built to assist object detection, we propose to pre-generate the relation map, and develop two-level shape-based constraints to gradually incorporate the relation information in the map to the detection model, costing no extra computations in inference time. 

\subsection{Multi-task Learning for Relation Modeling}
Multi-task learning (MTL) aims at training related tasks to leverage complementary information, providing each task with an inductive bias to trigger regularization effect between one another~\citep{caruana1997multitask}. One of the most successful examples where MTL is helpful for object detection is Mask RCNN~\citep{he2017mask}, which enhances the performance of object detection by jointly training an instance segmentation task. However, one of the main practical limitations is the requirement of mask labels, costing much more than bounding boxes.
Lots of researches in the medical domain have also corroborated the success of harnessing the relatedness to simultaneously improve performance of related tasks such as pancreas localization and segmentation~\citep{roth2018spatial}, and surgical tool presence detection and phase recognition~\citep{jin2020multi}. However, these frameworks also suffer from the above problems that each task requires its manual annotation.

To cope with aforesaid limitation and benefit from the prominent advantage of MTL, many attempts to automatically generate annotations for the specially designed auxiliary tasks have successfully made it~\citep{wang2020deep,lee2020structure,chen2020multi}.
For instance, \cite{wang2020deep} \rev{design} a geometry-aware tubular structure segmentation method. A distance transform mechanism is performed to generate a distance map as the auxiliary task for combining tubular structural cues to boost segmentation performance.
\cite{lee2020structure} \rev{present} a structure boundary preserving segmentation network, in which a boundary keypoint selection algorithm is proposed to automatically estimate the structural boundary of the target object for improving the segmentation results.
\cite{chen2020multi} \rev{build} a multi-task model to simultaneously detect shadow regions, shadow edges, and shadow count by leveraging their complementary information. 
Inspired by this stream of works, we propose to leverage the shape information of landmarks by automatically pre-generating the relation heatmaps, breaking the limitation of laborious annotations.  

\section{Method}
\begin{figure*}[t]
    \includegraphics[width=\textwidth]{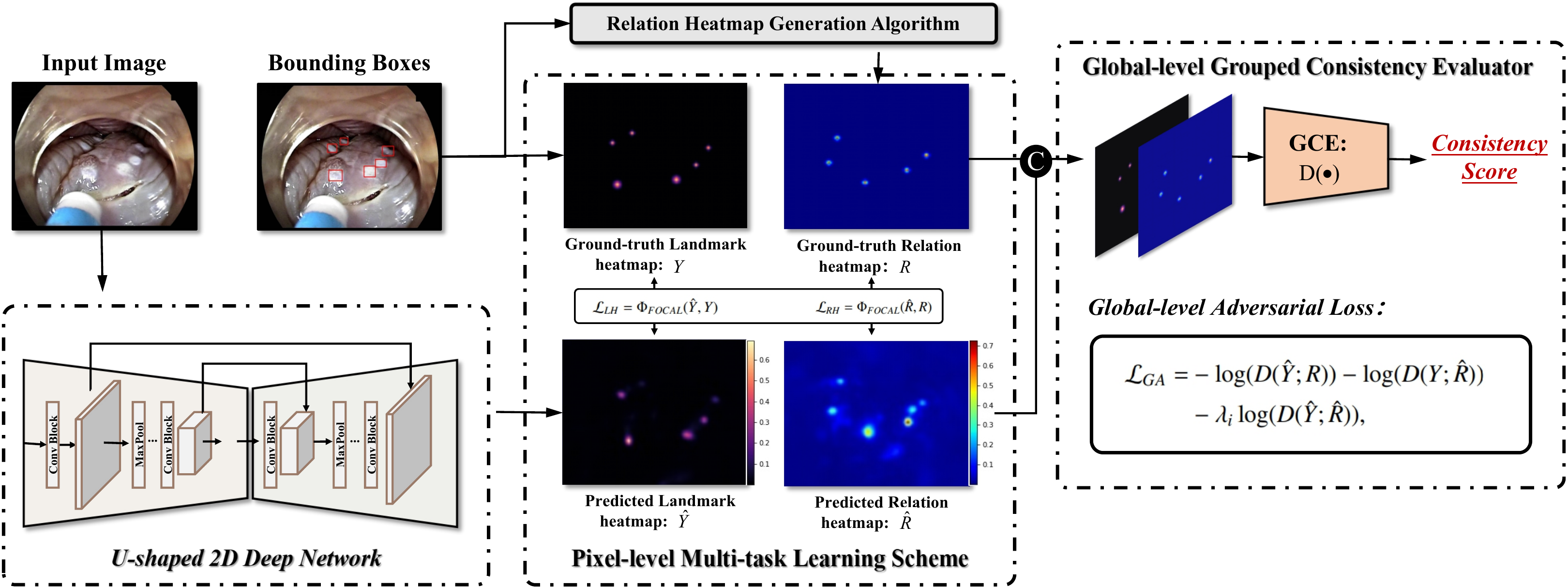}
    \caption{Overview of our proposed shape-aware relation network for real-time landmark detection in ESD surgery. Relation heatmap generation algorithm is designed to produce the relation heatmaps that well represent the prior knowledge of landmarks. Two complementary regularization schemes with pixel-level multi-task learning and global-level grouped consistency evaluator are developed, to progressively incorporate the relation prior to improve the detection model learning. ``C" denotes the channel-wise concatenate.
    }
    \label{framework}
\end{figure*}
An overview of our proposed shape-aware relation network for detecting ESD landmarks is illustrated in Fig.~\ref{framework}.
We first introduce our relation heatmap generation algorithm to produce relation heatmap that best represents the spatial correspondences of landmarks in ESD, to ensure the usefulness of relation detection.
Next, we present two complementary pixel-level and global-level regularization schemes aiming to integrate the spatial relation prior to facilitating the model learning.
Then, we show the training objectives for the relation learning in the detection model.
The training procedure and inference details are described in the end.

\subsection{Relation Heatmap Generation Algorithm}
\label{sec:generation-algorithm}
\begin{figure}[t]
    \includegraphics[width=\linewidth]{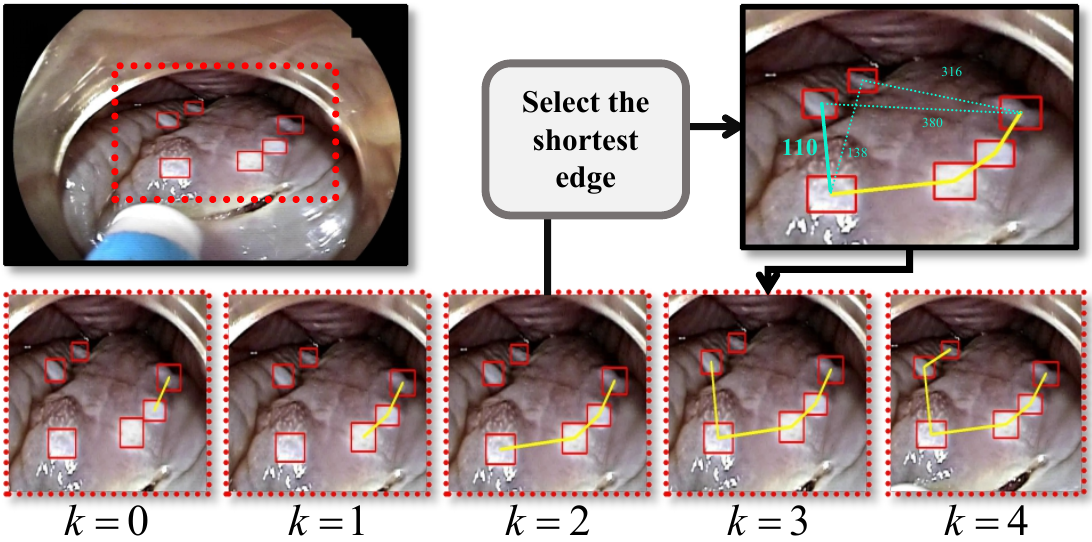}
    \caption{Iterative generation algorithm. Given six landmarks appearing in the example image, five iterations are performed for generating the relation keypoint heatmap for this image.}
    \label{fig:update}
\end{figure}

Effectively leveraging the inherent knowledge of the curve shape existing in the landmarks is essential.
However, selecting the correct cues lying at the curve is challenging, as the semantic features of the underlying curve are highly similar to the features of other regions (see the example in the top-left corner of Fig.~\ref{fig:update}).
We therefore devise an iterative generation algorithm, to generate a relation keypoint heatmap based on the distance measurement of landmarks.
It can best fit the ground-truth curve region, and explicitly represent the geometric relation between landmarks.

Specifically, we denote the $n$ landmark keypoints set of each frame as $P^{1:n}=[x^{1:n}, y^{1:n}]$, \rev{and our goal is to generate a relation heatmap $R$ containing $n-1$ relation keypoints.}
We first figure out the spatial order of the given landmarks $P$, by estimating an edge set $S^{1:n-1}$ consisting of $n-1$ edges.
\rev{Here, the edge set is also regarded as the mathematical definition of curve shape.}
To achieve this, we initialize $S$ by one edge that connects the closest keypoints in $P$, i.e., the two points with the shortest Euclidean distance.
Then, to obtain a ``curve" and bypass forming a ``tree", we regularize each point to connect no more than the other two points.
In this regard, we iteratively update $S$ by only calculating the distance between the unpaired points with the two endpoints of the formed curve in the current iteration, \rev{till all points have been connected}.
Taking the calculation at iteration $k=2$ as an example (see the top-right corner in Fig.~\ref{fig:update}), there exist two keypoints that have not been composed as a pair.
We compute the Euclidean distance between them and two endpoints of the curve.
The shortest edge (length = 110 in the example) is archived into $S$, and the next calculation iteration is invoked.

We then convert the estimated edge set to the keypoint heatmap, given that keypoint is the more tractable supervision signal and is easier for model training.
To better represent the relation of each pair of two neighbor landmarks, we choose the middle point of each edge in $S$.
Each middle point is then represented with Gaussian-shaped kernel~\citep{law2018cornernet} at its location.
After conversation of $n-1$ middle points, we can finally obtain a relation keypoint heatmap \rev{$R$}.

\subsection{Pixel-level Relation Regularization via Multi-tasking}
\label{sec:relation-network}
Pixel-wise estimation of the generated relation keypoint heatmap can encourage the model to gain the knowledge of spatial relation of landmarks.
We propose to form a multi-task learning paradigm jointly with the main task of landmark detection to perform this pixel-level relation constraint.

Given an input image $I \in \mathbb{R}^{W \times H\times 3}$ with width $W$ and height $H$, we employ a U-shaped 2D deep network~\citep{zhou2019objects} to extract representative high-level features, and output a convolutional feature map $\mathbf{F}$.
For conducting multi-tasks of detecting landmarks and their relations, the network construction splits into four branches. 
The first two branches contribute to the landmark detection by feeding the extracted feature map $\mathbf{F}$ into respective $3$$\times$$3$ convolutional layer, a ReLU activation function, and a $1$$\times$$1$ convolutional layer, outputting the keypoint heatmap $\hat{Y} \in \mathbb{R}^{\frac{W}{d} \times \frac{H}{d} \times 1}$ to indicate the landmark position, and the size value $\hat{S} \in \mathbb{R}^{\frac{W}{d} \times \frac{H}{d} \times 2}$, where $d$ is the downsample rate of prediction and empirically set as $4$.
To obtain more precise landmark position when mapping $\hat{Y}$ to the original size ($W \times H$),
the third branch is designed to compute offset value $\hat{O} \in \mathbb{R}^{\frac{W}{d} \times \frac{H}{d} \times 2}$, following ~\citep{zhou2019objects}. 
Here, we employ the same architecture as the previous two branches.

To leverage the prior knowledge of the relationship between the landmarks, we design the fourth branch targeting at detecting relation keypoint heatmap.
The ground truth for network training is from our devised automatic generation algorithm.
Without relying on any additional labeling, the shape-aware relation can be embedded in the model by optimizing the outputs towards our generated relation heatmap.
In practice, we also employ the same stack of layers (including $3$$\times$$3$ convolutional layer, a ReLU function, and a $1$$\times$$1$ convolutional layer) to form the fourth branch, and perform on $\mathbf{F}$ to predict the relation heatmap $\hat{R}$.
To this end, we form a multi-tasking model $M(\cdot)$ to simultaneously detect landmarks and their spatial relation.

\subsection{Global-level Relation Regularization via Grouped Consistency Evaluator}
Multi-task learning regularizes the model at the pixel level, by introducing the auxiliary branch to identify whether each pixel locates in the relation keypoint position or not.
To further leverage the relation knowledge, we can also regularize the model in a global view of the heatmap.
Given the high correlation in the keypoint position between the two kinds of maps, we can measure a consistency score after observing the whole heatmap.
A pair of ground-truth landmark and relation heatmaps should show higher consistency than the pair that any of them is replaced by the model predicted one.
In this regard, we propose a novel consistency-based discrimination network, named as Grouped Consistency Evaluator (GCE).
It then can provide the global-level relation constraint to the detection network via adversarial learning.

We design GCE based on VGG11\citep{2014Very} by adding an Instance Normalization layer, and a LeakyReLU function following each $3$$\times$$3$ convolutional layer.
In global, it receives a pair of landmark heatmap and relation heatmap as input and output a measurement score with the range from  0 to 1.
Each of them could be the predictions from our multi-task detection model ($\hat{Y}$ / $\hat{R}$), or the ground-truths ($Y$ / $R$, recall Section~\ref{sec:generation-algorithm} for the ground-truth generation of the relation heatmap).
The two maps are first concatenated across the channel-wise and then fed into GCE to get a score of their consistency. Thus, we have four types of heatmap pairs.
Only given a pair of the ground-truth landmark heatmap and relation heatmap, GCE shall provide a high evaluation score.
Otherwise, with the predicted landmark or relation heatmap, the output evaluation score shall be low.
In this regard, we train the evaluator through the following objective:
\begin{equation}
\begin{aligned}
    \mathcal{L}_{GCE} = &-\log(D(Y;R))-\lambda_{f}\log(1-D(\hat{Y};R))\\
    &-\lambda_{f}\log(1-D(Y;\hat{R}))-\lambda_{f}\log(1-D(\hat{Y};\hat{R})),\\
\end{aligned}
\end{equation}
where $D(\cdot)$ denotes the GCE function that projects the input relation heatmap and landmark heatmap onto a consistency-based evaluation score. 
Aiming at balancing the weight of less-consistent samples, we scale the three parts of negative loss by a constant $\lambda_{f}$, and $\lambda_{f}$ is set as $1/3$ in all experiments.

\subsection{Objective Functions for Training Detection Network}
\label{sec:detection-network}
To progressively incorporate the relation constraint in the detection network during the training procedure, we devise two types of loss functions to optimize the detection network.
The two losses respectively provide the pixel-level regularization via multi-task learning, and global-level regularization via adversary learning with GCE.

The first one is a multi-task detection loss $\mathcal{L}_{Mul}$ to optimize the model towards the ground-truths:
\begin{equation}
\begin{gathered}
\label{eq:mul}
\mathcal{L}_{Mul} = \mathcal{L}_{LH}+\alpha_{s}\mathcal{L}_{LS}+\alpha_{o}\mathcal{L}_{LO}+\alpha_{r}\mathcal{L}_{RH}, \\
\text{where} ~~ \mathcal{L}_{LH} = \Phi_{FOCAL}(\hat{Y},Y), ~~
\mathcal{L}_{LS} = \Phi_{\text{L1}}(\hat{S},S),\\
\mathcal{L}_{LO} = \Phi_{\text{L1}}(\hat{O},O), ~~
\mathcal{L}_{RH}=\Phi_{FOCAL}(\hat{R},R).
\end{gathered}
\end{equation}
$ \mathcal{L}_{LH}$, $\mathcal{L}_{LS}$, $\mathcal{L}_{LO}$ are the losses for the task of \textbf{l}andmark detection, including the  keypoint \textbf{h}eatmap prediction, and \textbf{s}ize and \textbf{o}ffset regression.
$\mathcal{L}_{RH}$ denotes the extensional loss for the auxiliary task of \textbf{r}elation keypoint \textbf{h}eatmap prediction.
We employ the pixel-wise logistic regression with the focal loss $\Phi_{FOCAL}$ to optimize heatmap~\citep{lin2017focal}: $\Phi_{FOCAL}(p_m) = -(1-p_m)^\gamma\log(p_m),$
where $p_m$ denotes the prediction probability of each pixel in the predicted heatmap. $\gamma$ is used to control the weight between negatives and positives, set as 2 following~\citep{lin2017focal}.
We use L1 loss $\Phi_{\text{L1}}$ for size and offset regression.
$\alpha_{s}$, $\alpha_{o}$, and $\alpha_{r}$ are the weights to balance the losses.
In all experiments, we empirically set $\alpha_{s}=0.1$, $\alpha_{o}=0.1$, and $\alpha_{r}=1$.

The other is a global-level adversarial loss $\mathcal{L}_{GA}$ to restrain landmark detection network, to let the multi-task detection model $M(\cdot)$ compete with the grouped consistency evaluator (GCE) $D(\cdot)$.
In other words, this objective is to increase the consistency of the predicted heatmaps from $M(\cdot)$, trying to fool $D(\cdot)$ by outputting the ground-truth-like heatmaps.
Our global-level adversarial loss is defined as:
\begin{equation}
\begin{aligned}
    \mathcal{L}_{GA} = &-\log(D(\hat{Y};R))-\log(D(Y;\hat{R}))\\
    &-\lambda_i\log(D(\hat{Y};\hat{R})),\\
\end{aligned}
\end{equation}
where $\lambda_i$ is also a hyperparameter used to weight the different loss items, since two predicted maps are more likely to be inconsistent compared to the ones containing a ground-truth map. Here we set $\lambda_i = 0.1$ in all experiments.

The overall loss for training detection model is a hybrid supervised loss ($\mathcal{L}_{DET}$), which is defined as:
\begin{equation}
\label{eq:overall}
\mathcal{L}_{DET} = \mathcal{L}_{Mul}+\alpha_{e}\mathcal{L}_{GA}. 
\end{equation}
$\alpha_{e}$ is used to balance the weight between two parts of supervised loss and set to 0.1 if not specially mentioned.
With two-level constraints in $\mathcal{L}_{Mul}$ and $\mathcal{L}_{GA}$, our detection model can sufficiently leverage the shape-aware relation existed in landmarks, and hence improve the detection performance.

\subsection{Training and Inference Procedure}
\paragraph{\textbf{Training Procedure}}
To sufficiently take advantage of the geometric relation between the landmarks, it is crucial to carefully design the training procedure. 
In practice, we exploit a three-step strategy to train our network:
(i) At first, we train the detection network for pure landmark detection. Here, we initialize the parameters of the network with a pre-trained model on COCO dataset~\citep{2014Microsoft} to avoid overfitting. 
(ii) Then we jointly train the detection network with the auxiliary task of relation heatmap prediction with Eq.~\ref{eq:mul}, i.e., the multi-tasking loss $\mathcal{L}_{Mul}$.
(iii) We next include the grouped consistency evaluator (GCE) and train the detection network in the adversarial manner towards Eq.~\ref{eq:overall}, i.e., the overall loss function $\mathcal{L}_{DET}$.
We find that all the loss terms tend to show a large fluctuation. Therefore, we reduce the training frequency of the evaluator in this stage. In other words, we train the evaluator every three times as a training detection part. 

\paragraph{\textbf{Efficient Inference}}
Compared with feature aggregation methods~\citep{xu2019spatial,lin2020dual}, another advantage of our proposed relation regularization is that, it alleviates to increase high calculation cost in the inference time.
For pixel-level regularization with multi-task learning, we remove the branch of predicting relation keypoint and generate the predicted bounding box by using original branches for object detection task ($\hat{Y}, \hat{S}, \hat{O}$). 
When predicting landmark heatmap $\hat{Y}$, we detect all responses whose value is greater or equal to its 8-connected neighbors.
For each peak $\hat{p}_i$, we adopt the keypoint value $\hat{Y}_{\hat{x}_i,\hat{y}_i}$ (i.e., prediction probability in the keypoint location ($\hat{x}_i,\hat{y}_i$)) as its detection confidence score, and keep the top 20 confident peaks.
We utilize the size value $\hat{S}_{\hat{x}_i,\hat{y}_i}$ as its width and height, deducing the ultimate outcome. 
Moreover, GCE is not required in the inference, therefore introduces no extra computational cost.
\section{Experiments}
\subsection{Dataset}
\begin{table}[t]
    \caption{Numbers of ESD frames and bounding box (Bbox) annotations in our dataset. ``S, M, L" respectively denote bounding box of small, middle, large size, according to the standard of COCO dataset. For each fold ($D_1-D_5$), we show the numbers of raw frames and Bbox for validation part. 
    }\label{table:dataset}
    \centering
    \renewcommand\arraystretch{1.2}
    \begin{tabular}{c|ccccc}
    \hline
    \hline
    Data & $Bbox_{S}$ & $Bbox_{M}$  & $Bbox_{L}$ & $Bbox_{ALL}$ & $Frame$ \\
    \hline
    $D_1$ & 26 & 1287 & 797 & 2110 & 643 \\
    $D_2$ & 36 & 1203 & 842 & 2081 & 838 \\
    $D_3$ & 1 & 920 & 1165 & 2086 & 747 \\
    $D_4$ & 4 & 888 & 1774 & 2666 & 774 \\
    $D_5$ & 37 & 965 & 916 & 1918 & 695 \\
    \hline
    $D_{ALL}$ & 104 & 5263 & 5494 & 10861 & 3697 \\
    \hline
    \end{tabular}
    
\end{table}

We newly collect a large in-house dataset of ESD surgery to validate our method, given that to the best of our knowledge, there is no publicly available ESD surgery dataset for this essential task of landmark detection.
The collected data consists of 11 ESD surgery procedures from several clinical centers in Xiamen ZhongShan Hospital.
The video is recorded by endoscope of FUJIFILM EG-580RD/EG-L580RD7 with 60 fps and resolution of $1920\times1080$.
Each video records all stages of ESD surgery.
As our work focuses on one of the most determinant stages \emph{circumferential cutting}, we use the video segments recording this stage in experiments.
We downsample the videos from 60 fps to 5 fps for reducing data redundancy, attaining $3,697$ frames in total.
\rev{Three experienced surgeons are involved in the ground truth labelling, i.e., bounding box of the landmark, and use a public annotation tool, Colabeler. 
Two junior surgeons help the first-round annotations with one responsible for 6 videos and the other one responsible for the rest videos.
One senior surgeon helps with the second-round check to ensure the accuracy of the annotations.}
With the bounding box, we further obtain the training targets of different branches in our multi-task detection model.
We use the Gaussian kernel to smooth each annotated boundary box around its center point to obtain the ground truth of landmark keypoint heatmap $Y$.
Meanwhile, we use the width and height of each annotated bounding box and pack them into a vector as the ground truth of landmark size $S$. 
The ground truth of relation keypoint heatmap $R$ is generated by our designed algorithm in Section~\ref{sec:generation-algorithm}.
For a comprehensive and convincing validation, we make the best use of annotated data and perform a 5-fold cross-validation on the dataset, with the detailed statistics of each fold shown in Table~\ref{table:dataset}.
Specifically, we split the fold in video-wise, obtaining folds $D_1-D_5$.
As each video contains different numbers of frames, the amount of training and validation frames (the numbers shown in Table~\ref{table:dataset}) in each fold is slightly different.
However, we do our best in data division, resulting in the nearly closed frame numbers in different folds.

\begin{table*}[ht]
    \caption{Comparison of landmark detection with different approaches with 5-fold cross-validation. For $AP$ and  $AP_{50}$, the results of each fold and the average results of all folds are presents in $D_{1-5}$ and $D_{ALL}$, respectively. For $Recall$, only average results of all folds are listed given the space limitation.}
    \label{exp}
\centering
\renewcommand\arraystretch{1}
\begin{tabular}{ccccccccccccccc}
\hline
\multirow{2}{*}{Method}&Speed&$Recall$
&\multicolumn   {6}{c}{$AP$~(\%)}
&\multicolumn{6}{c}{$AP_{50}$~(\%)}\\
\cline{4-15}
& (ms) & (\%) & $D_1$ & $D_2$& $D_3$& $D_4$& $D_5$& $D_{ALL}$
& $D_1$ & $D_2$& $D_3$& $D_4$& $D_5$& $D_{ALL}$\\
\hline
YOLOv3-tiny & 47& 60.0&
10.7& 16.1& 9.5& 15.1& 13.6& 12.8& 36.3& 48.5& 30.3& 53.4& 37.6& 42.0\\
YOLOv3      &70 &69.6& 
\bf{19.4}&19.6& 13.3& 28.9& 17.1& 18.2& 52.7& 56.0& 40.8& 68.4& 42.2& 51.7\\
YOLOv3-spp  &78 & 64.8 &
15.0& 18.8& 16.8& 26.1& 20.4& 19.3& 41.0& 53.4& 45.6& 67.2& 47.8& 51.1\\
\hline
EfficientDet-D0 &68 & 70.8&
13.4& 18.0& 12.9& 25.4& 20.8& 17.9& 37.2& 47.4& 35.6& 62.5& 45.2& 45.8\\
EfficientDet-D1 &87 & 71.4&
15.3& 17.2& 15.2& 26.3& 22.8& 18.8& 43.3& 46.5& 37.1& 66.4& 51.7& 48.2\\
EfficientDet-D2 &90& 73.4&
17.8& 15.6& 12.6& 30.0& 22.7& 19.2& 45.9& 46.3& 32.9& 69.0& 50.9& 48.8\\
\hline
CenterNet-Res18&25&62.8&
14.2& 17.6& 15.4& 23.4& 21.7& 18.3& 39.8& 49.1& 38.7& 62.5& 49.0& 48.4\\
CenterNet-DLA34 &27&70.9 
&17.2& 18.9& 19.2& 27.9& 23.8& 20.8& 46.4& 52.7& 51.3& 71.7& 54.9& 54.6\\
\hline
Ours-Res18& 25& 71.3&
18.3&20.0& 21.3& 26.6& 26.4& 22.0& 48.6& 55.5& 53.2& 69.4& \bf{59.0}& 57.0\\
Ours-DLA34 & 27 & \bf{74.1} &
19.3& \bf{21.4}& \bf{23.3}& \bf{30.7}& \bf{27.1}& \bf{22.7}& \bf{53.6}& \bf{57.3}& \bf{55.6}& \bf{72.4}& 58.0& \bf{58.1}\\
\hline
\end{tabular}
\label{tab:results}
\end{table*}
\begin{table*}[ht]
    \caption{Analytical ablation comparison of landmark detection with shape-aware regularization at different levels. ``Pixel", ``Global" denote pixel-level and global-level regularization, respectively. Values in the subscript denote the improvement compared to the baseline result without regularization.}\label{exp}
\centering
\renewcommand\arraystretch{1.2}
\setlength{\tabcolsep}{1.1mm}{
\begin{tabular}{cccccccccccccc}
\hline
\multirow{2}{*}{Pixel}&\multirow{2}{*}{Global}&\multicolumn{6}{c}{$AP$~(\%)}&\multicolumn{6}{c}{$AP_{50}$~(\%)}\\
\cline{3-14}
&& $D_1$ & $D_2$& $D_3$& $D_4$& $D_5$& $D_{ALL}$
& $D_1$ & $D_2$& $D_3$& $D_4$& $D_5$& $D_{ALL}$\\
\hline
 & & $17.2$ & $18.9$& $19.2$& $27.9$& $23.8$& $20.8$& $46.4$& $52.7$& $51.3$& $71.7$& $54.9$& $54.6$\\
\checkmark & & $19.3_{+2.1}$& $20.3_{+1.4}$& $19.6_{+0.4}$& $29.6_{+1.7}$& $26.3_{+2.5}$& $22.7_{+1.9}$& $48.7_{+2.4}$& $53.3_{+0.6}$& $48.3_{-3.0}$& $70.1_{-1.6}$& $56.6_{+1.7}$& $55.6_{+0.9}$\\
& \checkmark& $18.0_{+0.8}$& $20.4_{+1.4}$& $21.6_{+2.3}$& $28.8_{+0.9}$& $25.0_{+1.2}$& $21.1_{+0.3}$& $51.3_{+4.9}$& $54.7_{+2.0}$& $52.3_{+1.0}$& $69.0_{-2.7}$& $55.2_{+0.4}$& $55.0_{+0.4}$\\
\checkmark & \checkmark& $19.3_{+2.1}$& $21.4_{+2.5}$& $23.3_{+4.1}$& $30.7_{+2.8}$& $27.1_{+3.3}$& $22.7_{+1.9}$& $53.6_{+7.3}$& $57.3_{+4.6}$& $55.6_{+4.3}$& $72.4_{+0.7}$& $58.0_{+3.2}$& $58.1_{+3.5}$\\
\hline
\end{tabular}}
\label{tab:ablation}
\end{table*}
\subsection{Evaluation Metrics}
We adopt two widely-used metrics in object detection task as the quantitative measurement to evaluate our method: 
Average Precision ($AP$) that is under the varying IoU thresholds~($0.5:0.95:0.05$); and Average Precision ($AP_{50}$) at IoU threshold as 0.5. Readers can refer to \citep{law2018cornernet,zhou2019objects} for more detailed definitions. 
In our ESD landmark detection task, we alongside present the value of $Recall$, since this criterion shows the percentage of successfully detected landmarks.
Therefore, it can intuitively reflect the degree of alerts for surgeons by the detection model. 
Here we set the confidence threshold as $0.1$ and $IoU$ threshold as $0.5$ when calculating $Recall$, which is also a common setting in practice~\citep{law2018cornernet,zhou2019objects}.
In the following comparison, we list the performance of each fold, also show the average results of all five folds. 
We only show the overall result of $Recall$ considering the space limitation and aesthetic.
Moreover, we present the inference time per image, including the time for pre-processing, forward propagation and post-processing.
The result is the average value of all the inference images.

\subsection{Implementation Details}
We resize the frames from the original resolution of $1920 \! \times \! 1080$ to $512 \!  \times \!  512$ to save memory and reduce network parameters.
A series of augmentations is randomly used to increase the data diversity including horizon flip, rescaling~(between 0.6 and 1.4), shifting~(between 0.6 and 1.4) and color jittering. 
$128\times128$ cropping is then performed on the augmented image to further enlarge the training data.
We optimize the overall objective using Adam optimizer with initial learning rate of 0.0005 and batch size of 16.
We implement our framework based on PyTorch with 2 NVIDIA Titan RTX GPUs.
We totally train the network of 150 epochs with 50 epochs in each training step, which takes around 4 hours for training the entire framework.
To verify the general efficacy of exploring shape-aware relation between landmarks in our method, we utilize two network settings, i.e., developing the U-shaped 2D deep network for feature extraction based on two different backbones, \rev{including 18-layers ResNet (Res18)~\citep{he2016deep} and DLA34~\citep{yu2018deep}}.
The network architecture development follows~\cite{zhou2019objects}.

\begin{figure*}[ht]
    \includegraphics[width=\textwidth]{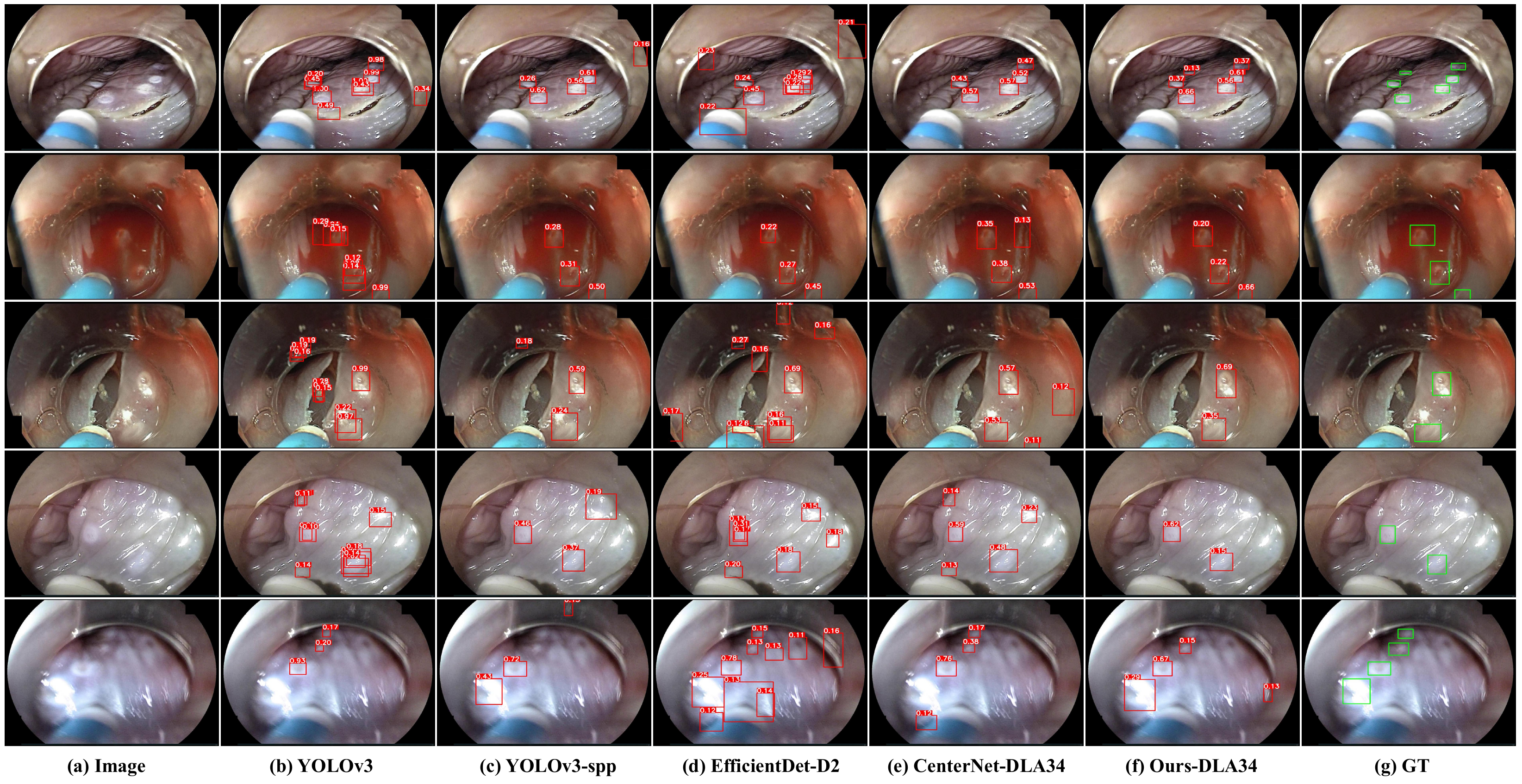}
    \caption{Visual comparison of different approaches on the challenging cases.  From left to right: (a) input images, (b-e) results of other compared methods, (f) result of our model with backbone of DLA34, (g) ground-truths (GT). We present the landmarks whose prediction confidences are larger than 0.1, and detailed confidence scores are illustrated in the top-left corner. Zoom in for better visualization.}
    \label{Result}
\end{figure*}
\subsection{Experimental Setup and Comparison}
 To evaluate our proposed method, considering that there are no existing works on this task, we compare it with several well-known detectors with real-time efficacy in the natural computer vision area.
Specifically, we choose the families of YOLOv3, EfficientDet, CenterNet as our compared frameworks: 
(i) YOLOv3 is the most applicable one-stage detection framework in the industry thanks to its critically acclaimed flexibility. 
The family of YOLOv3 contains many versions and we use the latest version for comparison, including YOLOv3-tiny, YOLOv3, YOLOv3-spp~\citep{glenn_jocher_2020_4308573}.
(ii) EfficientDet~\citep{Tan_2020_CVPR} is the latest state-of-the-art for real-time detection.
It has variations with eight different complexity levels of architecture (D0 to D7), and we compare with the variants of the first three levels (D0 to D2) considering the computational efficiency. 
(iii) CenterNet~\citep{zhou2019objects} is the most advanced anchor-free framework with real-time inference. It bypasses the anchor proposal generation in (i) and (ii), therefore, can largely alleviate the false positive detections.
Given that reducing false positive is a vital determinant to avoid providing confusing guidance for the surgeon, we adopt anchor-free architecture followed CenterNet, and compare it using two different network backbones, i.e., Res18 and DLA34.
For a fair comparison, we implement these methods with all of them pretrained on the COCO dataset and take the same input size of $512 \times 512$. 
In the non-maximum suppression operation of the anchor-based methods (YOLOv3 and EfficientDet), the overlap threshold is empirically set as 0.5.

\paragraph{\textbf{Quantitative Comparison}}
Table~\ref{tab:results} lists the quantitative results of different methods.
\rev{The family of YOLOv3 has the unsatisfactory $Recall$, demonstrating its in-satisfactory detection of those landmarks with similar semantics to background tissues. In addition, it's observed that they relatively perform worse on $AP$ than $AP_{50}$, indicating the less concrete regression of bounding boxes. While, due to the the shape-aware regularization that can reduce false detection and find more possible landmarks, our method has achieved much better results on all metrics.}
EfficientDet attains the better results of $Recall$ yet decreases in $AP$ and $AP_{50}$, due to massive false positive predictions left from anchor proposals.
The anchor-free method, CenterNet, shows improvement in $AP$ and $AP_{50}$ with DLA34 backbone. However, the results of $Recall$ are unsatisfactory, indicating the insufficient ability to detect landmarks.
By exploring shape-aware relation regularization, our method consistently outperforms CenterNet on all the metrics with two network backbones.
Interestingly, our method brings more obvious improvement on the backbone of Res18 ($AP$ raises $3.7\%$ and $AP_{50}$ raises $8.6\%$ compared with CenterNet). Moreover, it demonstrates that our shape-aware relation regularization is more beneficial for the simpler architecture to learn feature representation of landmarks.
Our method with DLA34 backbone peaks the best accuracy and retains a good inference speed, achieving $58.1\%$ $AP_{50}$ and $74.1\%$ $Recall$ with 27ms per frame. 
Besides the significant improvement in accuracy, the inference speed is also more than twice as fast as YOLOv3 and EfficientDet.

\begin{figure*}[ht]
\centering
    \includegraphics[width=\textwidth]{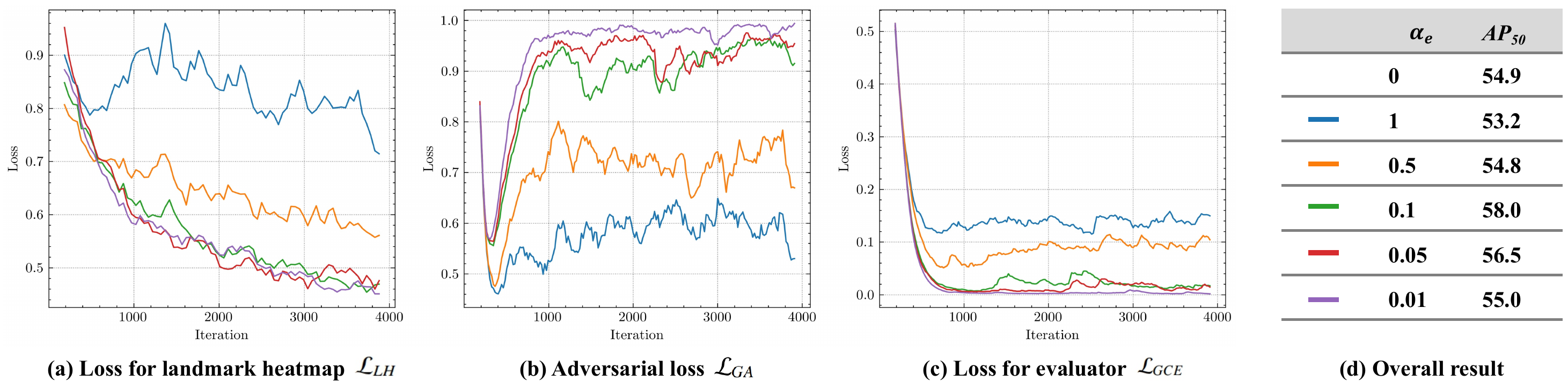}
    \caption{Comparison of experimental results at different values of $\alpha_e$ that controls the degree of global-level regularization. (a-c) Learning curves of different losses; horizontal axis denotes the learning iteration. (d) $AP_{50}$ results.}
    \label{fig:weight}
\end{figure*}
\paragraph{\textbf{Visual Comparison}}
We compare the visual results in the challenging cases shown in Fig.~\ref{Result}: the first row is a sample with multi-scale objects; the second and third rows present bleeding samples; the last two rows introduce samples with reflective noise and motion blur.
We present the detected bounding boxes of the landmarks whose detection confidence score are larger than $0.1$.
It is observed that by leveraging relation constraint, our framework can largely reduce false detections and recognize the landmarks with higher confidence.
Less false positives compared with CenterNet-DLA34 further verify that the improvements of our method are mainly derived from the shape-aware regularization.

\subsection{Analytical Ablation Study}
\paragraph{\textbf{Effectiveness of Key Components}}
We conduct extensive ablation experiments to validate the effectiveness of key components of our method, i.e., employing shape-aware relation regularization at different levels. 
In Table~\ref{tab:ablation}, we list four different configurations. 
The first row is the result of pure CenterNet-DLA34 as the baseline, and the rest three are the methods progressively integrating the pixel-level regularization via multi-task learning, global-level regularization via grouped consistency evaluator (GCE) and both-level regularizations. 
As our relation regularization does not bring the extra computational cost during inference, the inference time of different settings is the same. We therefore do not list the time for simplification.
It is observed that detection accuracy benefits from both levels of regularization: $AP_{50}$ raises $0.9\%$ using the pixel level and raises $0.4\%$ using the global level.
More importantly, the hybrid two-level regularization attains remarkable improvement of $3.5\%$ in $AP_{50}$, verifying the complementary advantage of the two-level regularizations.
We also see that individually utilizing pixel-level regularization yields better improvement compared with global-level regularization. 
The underlying reason is that pixel-level regularization can encourage the model to learn the geometric relation of landmarks in a more fine-grained wise, by explicitly optimizing the generated heatmap.

\paragraph{\textbf{Different Weights of Global-level Regularization}}
We further explore how the percentage of global-level regularization affect the detection performance, by varying the parameter $\alpha_e$ in Eq.~\ref{eq:overall}.
Specifically, we investigate a series of values, $\alpha_e=1,0.5,0.1,0.05,0.01,0$, where $\alpha_e=0$ means that the network is trained without GCE that provides the global-level regularization.
We illustrate detailed training loss curves of $\mathcal{L}_{Mul},\mathcal{L}_{GA}$, and $\mathcal{L}_{GCE}$ in Fig.~\ref{fig:weight} (a-c), and list results of $AP_{50}$ in Fig.~\ref{fig:weight} (d).
We adopt DLA34 as the network backbone and perform experiments on the fifth fold for saving time in this investigation.

We see that smaller values of $\alpha_e$ (0.01 or 0.05) show a slight influence on the final results.
When setting $\alpha_e$ to 0.1, the model achieves the highest $AP_{50}$. 
It is seen that although the adversarial loss is still large, it does provide effective regularization at the global level for landmark detection.
The underlying reason is that measuring the consistency is much simpler than generating consistent relation heatmap in our task. 
When increasing the value of $\alpha_e$ to 0.5 or 1, i.e., enlarging the impact of global-level regularization, performances are even inferior to the baseline with $\alpha_e = 0$.
In these configurations, we see that $\mathcal{L}_{Mul}$ is hard to decrease, indicating that too much constraint from adversary loss shall suppress the multi-task learning to predict landmarks, degrading the results eventually.

\paragraph{\textbf{GCE Interpretation by Visualizing Heatmaps}}
\begin{figure}[t]
    \centering
    \includegraphics[width=\linewidth]{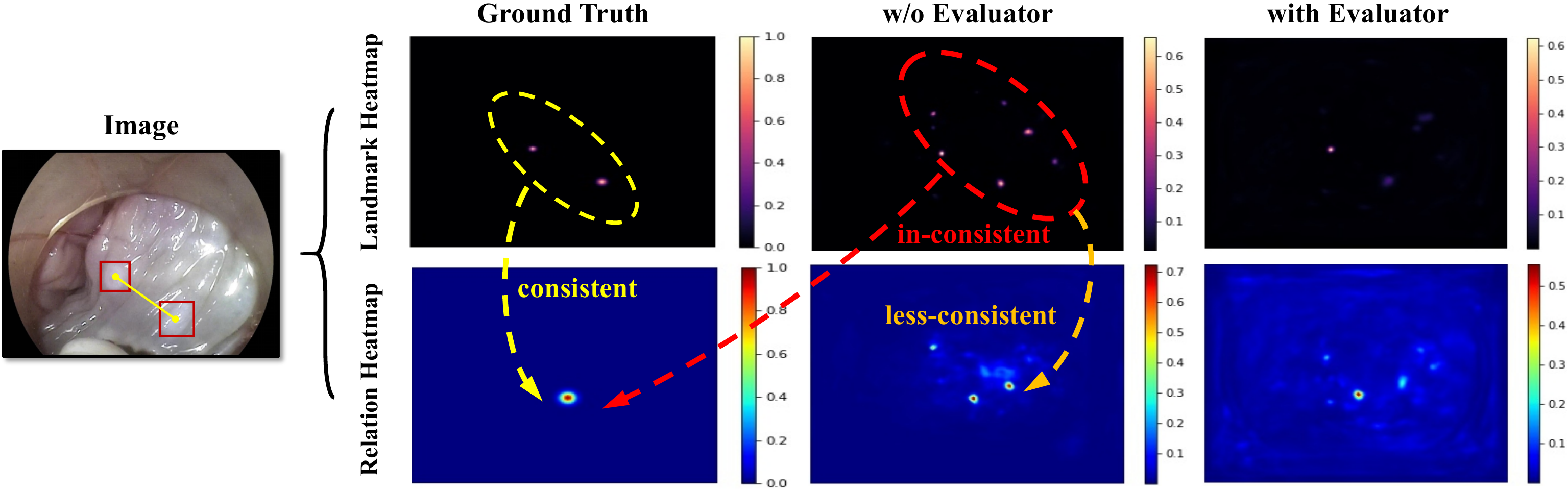}
    \caption{Illustration of heatmaps to intuitively interpret the insight of GCE design and how it works. From left to right: the input image, its ground-truths, predictions of the model without or with evaluator.}
    \label{Evaluator}
\end{figure}
In order to intuitively provide the insight of how grouped consistency evaluator (GCE) regularizes the model learning, we visualize the predicted heatmaps of models with or w/o GCE in Fig.~\ref{Evaluator}. \rev{As it shows, the global regularization can enhance the consistency of detected landmarks and their relation, therefore reducing the false detection of landmarks as well as relation by a large margin.}
We see that there exist several false detections in landmark heatmaps when do not utilize evaluator for global-wise constraint. Meanwhile, the estimated relation heatmap also contains false detection. The consistency of landmark geometric locations in the two predicted heatmaps has been destroyed. Additionally, the consistency between the predicted one and ground-truth is also inferior to that of the ground-truth pair.
Considering this phenomenon, we propose the GCE to regularize the model in a global view.
We can observe that, after equipped with GCE for network training, the false detected landmarks can be clearly removed.
The relation heatmap also shows better visualization results, in which only one mimic is detected with relatively low confidence.
More precise relation predictions shall in turn bring more beneficial regularization in landmark detection.

\begin{figure*}[!t]
\includegraphics[width=\textwidth]{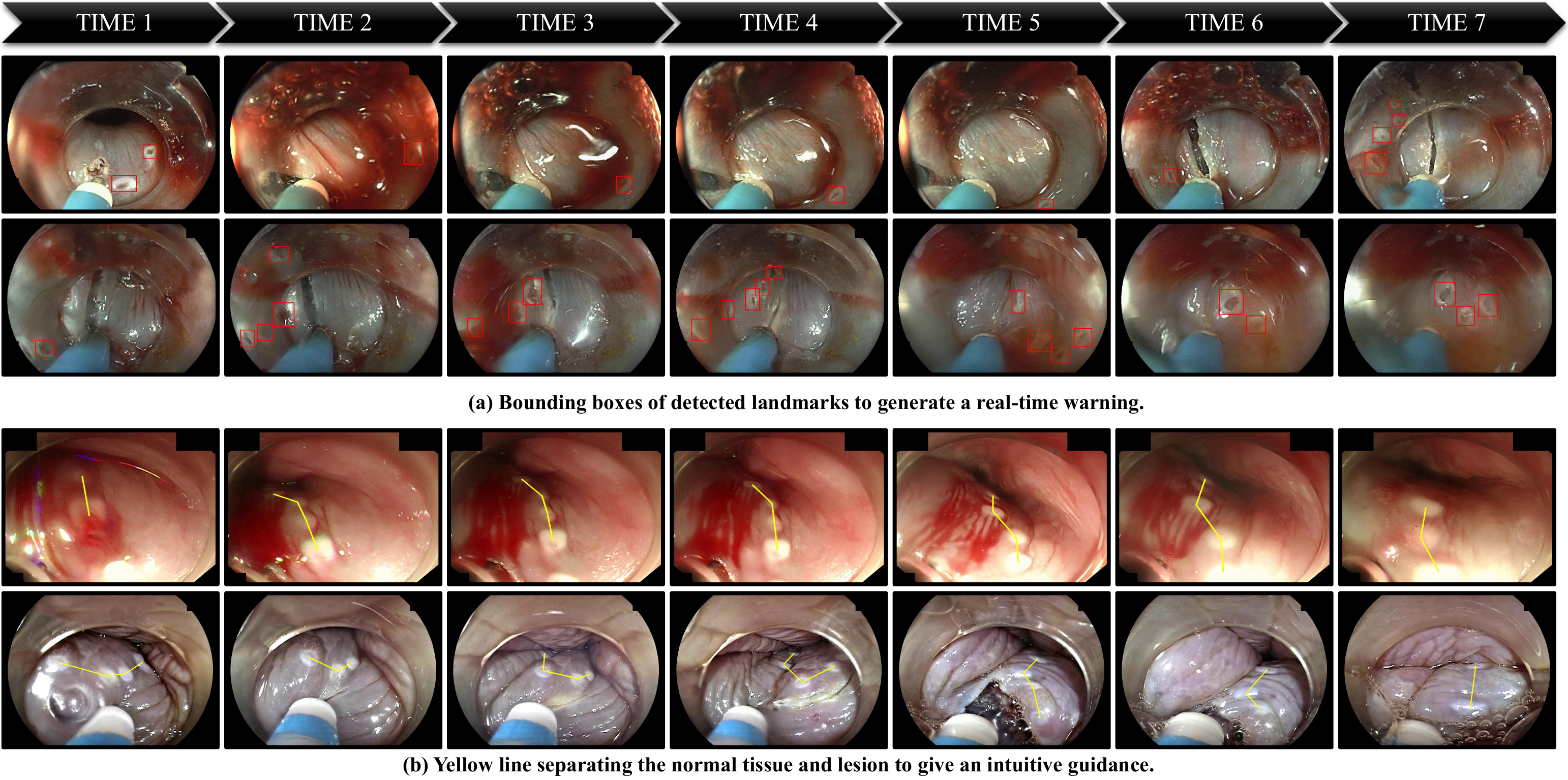}
\caption{Two potential downstream applications of our method in ESD surgery. Four typical video segments are presented with seven continuous frames in each segment. (a) For the complex surgical scenario, red bounding boxes of detected landmarks are illustrated to generate real-time warnings. (b) For the relatively clean surgical scene, yellow line estimating the boundary between normal tissue and lesion is presented, to give a more intuitive cutting guidance. 
} \label{fig:cas}
\end{figure*}
\subsection{Potential for Downstream Applications}
We further show the promising potential of our method on the crucial downstream applications in ESD surgery.
With the precise landmark detection and real-time efficacy, our method can provide two types of guidance signals to assist surgeons in practical ESD surgery.
Specifically, our method can generate the bounding boxes of landmarks and the boundary separating the tissue and lesion as shown in Fig.~\ref{fig:cas}, which supplies sufficient flexibility for surgeons' choice to confront different operation situations.
One common scenario is the landmarks covered by blood and nearly cannot be witnessed. 
In this situation, our method aims at sending out the warning signals to prevent the surgeons from cutting on landmarks (Fig.~\ref{fig:cas} (a)). 
We illustrate the bounding boxes of the detected landmarks and set a low confidence threshold of $0.1$, to show more probable landmarks.
Another scenario is that the whole surgical scene is relatively clean.
In this easier situation, our method aims at directly giving more intuitive guidance of cutting position (Fig.~\ref{fig:cas} (b)), which can support the decision-making, especially for the less experienced surgeons.
We increase the threshold to $0.2$ to solely keep landmarks detected with high confidence.
The boundary (visualized as yellow lines) is then produced by our designed generation algorithm in Section~\ref{sec:generation-algorithm} on the detected landmarks.

\section{Discussion}
Witnessing landmarks during the circumferential cutting stage is one of the most vital determinants for the successful resection in ESD surgery. 
Due to the challenging issues in the complicated endoscopic scene, it is extremely difficult for surgeons to notice some landmarks in time.
In this regard, developing the technology for real-time \emph{landmark detection} is highly demanded, which can improve the quality of ESD surgery and increase patient safety.
Nevertheless, scarce literature has paid attention to this crucial task.
In this paper, we take the first step for investigating this task, and develop a novel shape-aware relation network by wisely leveraging the prior knowledge in landmark locations in ESD surgery. 
We also carefully consider the computational cost when designing our method, as efficiency is an essential requirement in this task.
To validate the effectiveness of our method, we construct a real-world dataset of ESD surgery for esophageal cancer and conduct extensive experiments on it.
We further develop two downstream applications based on our precise detection, verifying the great potential of our method for clinical usage.

Landmark detection from the pure surgical scene in ESD surgery is very challenging. 
Fortunately, we observe a useful prior knowledge of ESD landmarks in terms of their geometric location.
As the landmarks are marked to indicate the boundary between the tissue and lesion, they have a high spatial relation, that is generally arranged in a curve shape.
Learning relations between objects has been proved beneficial for object detection tasks in the natural computer vision area.
However, most works focus on designing the complex modules embedded into the network to capture the relationship, such as non-local block~\citep{Deng_2019_ICCV} and graph convolutional layer~\citep{xu2019spatial}.
The whole network is self-contained and these modules need to be preserved during the inference, bringing the increase in processing time.
Instead, our method utilizes the shape-aware relation regularization by two-level auxiliary tasks, i.e., multi-task learning via introducing another branch and adversary learning via adding evaluator.
Both of them progressively benefit the model learning by only affecting the network training, and can be excluded and involve no extra computational cost in inference time.
In this regard, our method enjoys supplying online guidance to surgeons during the operation.

Surgery video is actually a form of sequential data, and effectively capturing the sequential dynamics has shown its successes in various surgical tasks, such as workflow recognition and tool presence detection in cholecystectomy procedure\rev{~\citep{jin2018sv,czempiel2020tecno,jin2021temporal} and instrument segmentation~\citep{zhao2021one}.}
However, in our work, we carefully consider the characteristic of ESD surgery and treat the video data as a static frame without using the temporal information.
The main determinant is the proportion of the \textit{hard frames} existing in different surgery procedures, i.e., the percentage of the aforementioned challenging cases, such as the scene occluded by blood or electric smoke, containing the complicated lighting changes, and motion blur.
In other surgeries, such as cholecystectomy, the \textit{hard frames} account for a small percentage, having little effect on the temporal continuity in video segment.
Therefore modeling temporal cues of video segment can benefit the feature representation learning for frames in this segment.
Instead, the \textit{hard frames} widely appear in ESD surgery. 
Simply applying temporal aggregation techniques, such as long short term memory (LSTM), tends to degrade performance due to much noise in each segment.
Additionally, using temporal aggregation shall lead to more computational and memory costs, harming the real-time inference efficacy.
Although this work does not use the temporal cues, our detected landmarks still demonstrate the satisfactory temporal consistency, as shown in Fig.~\ref{fig:cas}.
In future, we plan to design efficient and selective modules to leverage temporal cues of ESD surgery.
Ideally, such modules can avoid involving noise and only incorporate the informative frames.

\rev{Apart from the use of temporal information, another limitation is about the relation generation algorithm, that may not work well in some special cases. As said in Section~\ref{sec:generation-algorithm}, our relation set is updated according to the measurement of distance, while the distance in 2D vision can not always represent the true distance in the real world, which may lead to the wrong update order. In this regard, the relation generation algorithm sometimes is not able to provide suitable shape regularization. However, these cases cover little portion in the surgical images, since at most time, surgeons find the best view to perform cutting. In the future work, we shall explore a more convincing algorithm to calculate the relation, with the use of temporal information.}

Our proposed method has a significant clinical value by achieving accurate landmarks detection in ESD surgery.
The online efficacy (quite fast processing speed, i.e., 25-27 ms per frame) further widens the applicability of our method in clinical practice.
In this regard, our approach can help to improve ESD surgery quality, by providing context-aware decision support, generating real-time warning and guidance for surgeons.
It can also reduce the intra-operative adverse events, and further increase patient safety.
We establish two types of downstream applications to intuitively illustrate the clinical value of this work in real-world practice. 
Apart from these, our method is beneficial to many other surgical video analysis tasks in ESD surgery, such as segmentation of lesion region, trajectory prediction of the circumferential cutting.
In the future, we will explore the aforementioned promising directions starting with this work.

\rev{ESD surgery plays a vital role in removal of epithelial lesions. Our work aims to facilitate precise lesion removal at the stage of circumferential cutting by offering accurate landmark detection. Besides the task we explored, there are several other crucial computer-aided tasks for ESD surgery which are worth to be investigated in future, such as: (1) offering recommended position where the landmarks should be created, to ensure the complete removal of lesions; (2) computer-aided decision-making about standard depth during submucosal dissection, as it ensures the full removal of submucosal lesions; (3) automatic recognition of blood vessel that can assist surgeons to avoid cutting the vessels, enhancing the patients' safety. The computer-aided ESD surgery is a complicated project that requires many years of efforts to improve methodology and accumulate knowledge for the final accomplishment. We hope that our proposed AI powered computer-aided method can inspire other researchers to be dedicated to this valuable direction.}

\section{Conclusion}
In this paper, we propose a novel shape-aware relation network, to investigate an essential task of landmark detection in ESD surgery.
It is an efficient detector integrated with a complementary two-level regularization scheme to learn the inherent prior knowledge of spatial relation in landmarks.
To validate the advancement of our methods, we construct a large in-house dataset consisting of 11 entire ESD surgical videos. 
Experimental results demonstrate a great improvement of our method in landmarks detection on both accuracy and efficiency, compared to state-of-the-art detectors.
On top of this, we further build two downstream applications to demonstrate the promising potential of our method in clinical practice.

\section*{Acknowledgement.} 
This work was supported by the Fundamental Research Funds for the Central Universities (Grant No. 20720190012) \rev{and a grant from the Hong Kong Research Grants Council (No. PolyU 152035/17E)}.

\bibliographystyle{model2-names.bst}\biboptions{authoryear}
\bibliography{refs}
\end{document}